\documentstyle[11pt]{article}
\title{$b \rightarrow s(d) \gamma$ with a Vector-like Quark \\
	as Fourth-generation }
\author{L.T. Handoko\thanks{On leave of absence from Applied
	Physics Research and Development Center - Indonesian Institute of
	Sciences (PFT-LIPI), Serpong-INDONESIA.	E-mail address :
	handoko@theo.phys.sci.hiroshima-u.ac.jp} and T. Morozumi\thanks{
	E-mail address : morozumi@fusion.sci.hiroshima-u.ac.jp}}
\date{\small \it Department of Physics, Hiroshima University \\
	1-3-1 Kagamiyama, Higashi Hiroshima - 724, Japan}
\input{epsf}
\begin{document}
\setlength{\baselineskip}{24pt}

\maketitle
\begin{picture}(0,0)
\put(325,240){HUPD-9409}
\put(325,225){August, 1994}
\put(325,210){hep-ph/9409240}
\end{picture}
\vspace{-24pt}

\setlength{\baselineskip}{8mm}
\renewcommand{\thesubsection}{\Roman{subsection}}

\begin{abstract}
	We study the implications of a vector-like
	quark, especially a down-type vector-like quark in
	$b \rightarrow s(d) \gamma$ including the neutral Higgs in the
	calculation. After analyzing some possible constraints for the mixing,
	the contribution may be significant for the $b \rightarrow d \gamma$,
	but not for the $b \rightarrow s \gamma$. We also find that the FCNC
	among ordinary quarks vanishes for a special form of down-type quark
	mass matrix.
\end{abstract}

\subsection{\bf INTRODUCTION}

\hspace*{1em} Through study of the rare $B$ meson decays, we may learn some
clues about Yukawa-sector of the electroweak theory which has not been
understood well. In this meaning, it is important to test the various
extended quark sectors beyond the Standard Model (SM).
In this paper, as a possible minimum extension, we introduce a singlet
vector-like quark. In this Vector-like quark Model (VM), there are flavor
changing neutral currents (FCNC) among the ordinary quarks
and between the ordinary
quarks and the vector-like quarks. These two different types of FCNC, i.e., (1)
FCNC among the ordinary quarks and (2) FCNC between the ordinary and the vector-like
quarks can be tested separately. It can be easily understood that
$B^0 - \bar{B^0}$ mixing and $b \rightarrow s(d) l^+ l^-$ processes are
appropriate to test the case (1). This is because the FCNC
among the ordinary
quarks contributes to these processes in the tree-level. On the other hand,
in order to test the case (2), $b \rightarrow s(d) \gamma$ can be an
appropriate process since the FCNC between the ordinary quarks and the
vector-like quarks is comparable with the FCNC between the ordinary quarks.
The reason is, in the $b \rightarrow s(d) \gamma$ both of them occur in the one-loop level and
compete each other. On the other hand, in the $b \rightarrow s(d) l^+ l^- $
the FCNC between the ordinary quarks and the vector-like quarks contributes in the
one-loop level and is suppressed compared with the FCNC
among the ordinary quarks
which contributes in the tree level.

This paper is organized as follows. First we briefly describe the model.
Here we find that there is not only FCNC due to $Z$ exchange diagram, but
also FCNC due to neutral Higgs exchange diagrams. After computing the
$b \rightarrow s(d) \gamma$, we compare the main contributions, i.e.,
the charged current, the neutral current in the ordinary quark sector and
the neutral current between the ordinary quarks and the vector-like quarks.
In particular, we examine the dependence of the branching-ratio on the
vector-like quark mass.

\subsection{\bf THE MODEL}

\hspace*{1em} We study the Standard Model (SM) with an extended quark sector.
In addition to the three standard generations of quarks, we introduce one
down-type and one up-type vector-like quark,
\[
	{\tilde{\psi}_L}^i \equiv \left(
		\begin{array}{c}
			{\tilde{u}_L}^i \\
			{\tilde{d}_L}^i
		\end{array}
		\right) \: , \:
	{\tilde{u}_L}^4 \: , \: {\tilde{d}_L}^4 \: , \:
	{\tilde{u}_R}^{\alpha} \: , \: {\tilde{d}_R}^{\alpha}
\]
where $\tilde{u}^{\alpha}$ and $\tilde{d}^{\alpha}$ represent up-type and
down-type quarks in weak bases, with generation indices $\alpha = 1,2,3,4$
and $i = 1,2,3$. Throughout this paper, we use the following notations for
chirality, $L \equiv \frac{1}{2} (1 - \gamma_5)$ and
$R \equiv \frac{1}{2} (1 + \gamma_5)$.

In the previous works (ref. \cite{morozumi}, \cite{nir}, \cite{branco}),
the FCNC in the $Z$ sector has been studied already, but the FCNC in the
neutral-Higgs of the present model has been ignored. So let us explain briefly
how the FCNC comes out in the neutral Higgs sector. Here we consider one
Higgs doublet and denote them as below,
\begin{eqnarray}
	\phi & \equiv & \left(
			\begin{array}{c}
				\chi^+ \\
				\frac{1}{\sqrt{2}} \left( v + \phi^0 \right)
			\end{array}
			\right) , \nonumber \\
	\tilde{\phi} & \equiv & \left(
			\begin{array}{c}
				\frac{1}{\sqrt{2}} \left( v + {\phi^0}^{\ast}
				\right) \\
				{-}\chi^{-} \\
			\end{array}
			\right) ,  \\
	\phi^0 & \equiv & H + i\chi^0 . \nonumber
\end{eqnarray}
Then, the Lagrangian for Yukawa sector is,
\begin{eqnarray}
	{\cal L}_Y & = & -{f_d}^{i \alpha} {\bar{\tilde{\psi}}_L}^i \,
		\phi \, {\tilde{d}_R}^{\alpha}
	- {f_d}^{4 \alpha} {\bar{\tilde{d}}_L}^4 \,
		{\tilde{d}_R}^{\alpha} \, v^{\prime}
	- {f_u}^{i \alpha} {\bar{\tilde{\psi}}_L}^i \phi {\tilde{u}_R}^{\alpha}
	- {f_u}^{4 \alpha} {\bar{\tilde{u}}_L}^4 \,
		{\tilde{u}_R}^{\alpha} \, v^{\prime}
	+ h.c. \: . \label{eqn:higgs}
\end{eqnarray}
Here, ${f_d}^{i \alpha}$ and ${f_u}^{i \alpha}$ are the Yukawa couplings
which give the masses of fermions through the vacuum expectation value
of Higgs doublet as in the SM. The terms which are proportionated to
${f_d}^{4 \alpha}$ and ${f_u}^{4 \alpha}$ are bare mass terms. These
terms contain both diagonal couplings for vector-like quarks
($d^4$, $u^4$), and off-diagonal couplings between left handed $SU(2)$
singlet quarks (${d_L}^4$, ${u_L}^4$) and right handed singlet quarks
(${d_L}^i$, ${u_L}^i$). We diagonalize the Yukawa couplings as,
\begin{eqnarray}
	{f_d}^{i \alpha} \frac{v}{\sqrt{2}} \equiv
		\left( {V_L}^{\dagger} m_d V_R \right)^{i \alpha} & , &
	{f_d}^{4 \alpha} v^{\prime} \equiv
		\left( {V_L}^{\dagger} m_d V_R \right)^{4 \alpha} , \\
	{f_u}^{i \alpha} \frac{v}{\sqrt{2}} \equiv
		\left( {U_L}^{\dagger} m_u U_R \right)^{i \alpha} & , &
	{f_u}^{4 \alpha} v^{\prime} \equiv
		\left( {U_L}^{\dagger} m_u U_R \right)^{4 \alpha} ,
\end{eqnarray}
where $V$ and $U$ are $4 \times 4$ unitary matrices which relate the
weak-eigenstates to mass-eigenstates as,
\begin{eqnarray}
	{d_L}^{\beta} & \equiv & V^{\beta\alpha} {\tilde{d}_L}^{\alpha},
	\label{eqn:w1} \\
	{u_L}^{\beta} & \equiv & U^{\beta\alpha} {\tilde{u}_L}^{\alpha},
	\label{eqn:w2} \\
	V_{CKM}^{\alpha\beta} & \equiv & \sum_{i=1}^{3}
		U^{i\alpha} {V^{i\beta}}^{\ast} .
	\label{eqn:w3}
\end{eqnarray}
Substituting the above relations into the Lagrangian (\ref{eqn:higgs}),
the neutral Higgs sector becomes,
\begin{eqnarray}
	{\cal L}_{\phi^0} & = & -\frac{1}{v} {m_d}^{\beta} \,
		V^{\alpha i} {V^{\beta i}}^{\ast}
		{\bar{d}_L}^{\alpha} \, {{d}_R}^{\beta} \, \phi^0
	- \frac{1}{v} {m_u}^{\beta} \, U^{\alpha i} {U^{\beta i}}^{\ast}
		{\bar{u}_L}^{\alpha} \, {{u}_R}^{\beta} \, {\phi^0}^{\ast}
		+ h.c. \: .
	\label{eqn:nhs}
\end{eqnarray}
With the following definitions,
\begin{eqnarray}
	{z_d}^{\alpha\beta} & \equiv & \sum_{i=1}^{3} V^{\alpha i}
		{V^{\beta i}}^{\ast} = \delta^{\alpha\beta} -
		V^{\alpha 4} {V^{\beta 4}}^{\ast} ,
	\label{eqn:w4} \\
	{z_u}^{\alpha\beta} & \equiv & \sum_{i=1}^{3} U^{\alpha i}				{U^{\beta i}}^{\ast} = \delta^{\alpha\beta} -
		U^{\alpha 4} {U^{\beta 4}}^{\ast} ,
	\label{eqn:w5}
\end{eqnarray}
the neutral Higgs sector (eq. \ref{eqn:nhs}) can be written as,
\begin{eqnarray}
	{\cal L}_{\phi^0} & = & -\frac{1}{v} {m_d}^{\beta} \,
		{z_d}^{\alpha \beta}
		{\bar{d}_L}^{\alpha} \, {{d}_R}^{\beta} \, H
	- \frac{1}{v} {m_u}^{\beta} \, {z_u}^{\alpha \beta}
		{\bar{u}_L}^{\alpha} \, {{u}_R}^{\beta} \, H
	\nonumber \\
	&  & -\frac{i}{v} {m_d}^{\beta} \,
		{z_d}^{\alpha \beta}
		{\bar{d}_L}^{\alpha} \, {{d}_R}^{\beta} \, \chi^0
	+ \frac{i}{v} {m_u}^{\beta} \, {z_u}^{\alpha \beta}
		{\bar{u}_L}^{\alpha} \, {{u}_R}^{\beta} \, \chi^0
		+ h.c. \: .
\end{eqnarray}
It is apparent that if there are non-zero off-diagonal elements in
${z_d}^{\alpha \beta}$ and ${z_u}^{\alpha \beta}$, FCNCs arise in the neutral
Higgs sector.

About the gauge-fixing term, we employ the procedure of ref. \cite{grig},
\begin{eqnarray}
	{\cal L}_{GF} & \equiv & -\frac{1}{\xi} \left(f_W {f_W}^{\dag}
		+ {f_Z}^2 \right) ,
	\label{eqn:gf}
\end{eqnarray}
where,
\begin{eqnarray}
	f_W & \equiv & \partial_{\mu}{W^{\mu}}^-
		+ i e A_{\mu} {W^{\mu}}^- + i \xi M_W \chi^- ,  \\
	f_Z & \equiv & \frac{1}{\sqrt{2}} \left(\partial_{\mu}Z^{\mu}
		+ \xi M_Z \chi^0\right)  .
\end{eqnarray}
According to ref. \cite{grig}, the advantage of this gauge-fixing
is the absence of $W^{\pm} \chi^{\mp} A$ interaction in the full Lagrangian.
Therefore the number of diagrams is reduced without changing the propagator
of each field. For simplicity, we take $\xi = 1$ (Feynman gauge) along the
calculation.

Eventually, we can write the full Lagrangian for each sector as,
\begin{eqnarray}
	{\cal L} & = & {\cal L}_{W^{\pm}} + {\cal L}_{\chi^{\pm}} +
		{\cal L}_A + {\cal L}_Z + {\cal L}_H + {\cal L}_{\chi^0} ,
	\label{eqn:lagrangian}
\end{eqnarray}
where,
\begin{eqnarray}
	{\cal L}_{W^{\pm}} & = & \frac{g}{\sqrt{2}}
		V_{CKM}^{\alpha\beta} \bar{u}^{\alpha} \gamma^{\mu} \, L \,
		d^{\beta} \, W_{\mu}^+ + h.c. , \\
	{\cal L}_{\chi^{\pm}} & = & \frac{g}{\sqrt{2} M_W}
		V_{CKM}^{\alpha\beta}
		\bar{u}^{\alpha} \left( m_{u^{\alpha}} L -
		m_{d^{\beta}} R \right) d^{\beta} \chi^+ + h.c. , \\
	{\cal L}_A & = & \frac{e}{3} \left(
		2 \bar{u}^{\alpha} \gamma^{\mu} u^{\alpha}
		- \bar{d}^{\alpha} \gamma^{\mu} d^{\alpha}
		\right) \, A_{\mu} , \\
	{\cal L}_Z & = & \frac{g}{2\cos\theta_W}
		\left\{ \bar{u}^{\alpha} \gamma^{\mu} \left[\left(
		{z_u}^{\alpha\beta} - \frac{4}{3} \sin^2\theta_W
		\delta^{\alpha\beta}
		\right) L - \frac{4}{3} \sin^2\theta_W \delta^{\alpha\beta} R
		\right] u^{\beta} \right. \nonumber \\
	& &	+ \left. \bar{d}^{\alpha} \gamma^{\mu}
		\left[\left( \frac{2}{3} \sin^2\theta_W \delta^{\alpha\beta} -
		{z_d}^{\alpha\beta}\right) L + 							\frac{2}{3}\sin^2\theta_W \delta^{\alpha\beta}
		R\right] d^{\beta}\right\} \, Z_{\mu} , \\
	{\cal L}_H & = & \frac{-g}{2 M_W} \left[
		{z_u}^{\alpha\beta} \bar{u}^{\alpha}
		\left(m_{u^{\alpha}} L + m_{u^{\beta}} R\right) u^{\beta}
		\right. \nonumber \\
	& &	\left. \: \: \: \: \: \: \: \:
		+ {z_d}^{\alpha\beta} \bar{d}^{\alpha}
		\left(m_{d^{\alpha}} L + m_{d^{\beta}} R\right) d^{\beta}
		\right] \, H ,  \\
	{\cal L}_{\chi^0} & = & \frac{-ig}{2 M_W} \left[
		{z_u}^{\alpha\beta} \bar{u}^{\alpha}
		\left(m_{u^{\alpha}} L - m_{u^{\beta}} R\right) u^{\beta}
		\right. \nonumber \\
	& &	\left. \: \: \: \: \: \: \: \:
		- {z_d}^{\alpha\beta} \bar{d}^{\alpha}
		\left(m_{d^{\alpha}} L - m_{d^{\beta}} R\right) d^{\beta}
		\right] \, \chi^0 .
\end{eqnarray}
The terms which are proportionated to $m_{\alpha}/M_W$ in the
${\cal L}_H$ and ${\cal L}_{\chi^0}$ may give a significant contribution
due to the quarks in the internal line which have large mass $m_{\alpha}$.

\subsection{\bf$ b \rightarrow s(d) \gamma $ PROCESSES}
\label{sec:proc}

\hspace*{1em} We are now in position to evaluate the $b \rightarrow q \gamma$
($q = s, d$) in the model. Of course, it is also considerable to introduce an
up-type vector-like quark apart from a down-type vector-like quark, but
it is clear that in that case the contribution is exactly same with the
$SU(2)$ doublet Fourth-generation Model. This is because the new contribution
is coming from the addition of the up-type vector-like quark in the internal
line of $W^{\pm}$ and $\chi^{\pm}$ exchange diagrams. For that reason, we are
here only interested in the contribution of the down-type vector-like quark
where the new contribution is coming from the $Z$ and the neutral Higgs
exchange diagrams.

\begin{figure}[h]
	\epsfxsize=15cm
	\epsfysize=10cm
	\epsffile{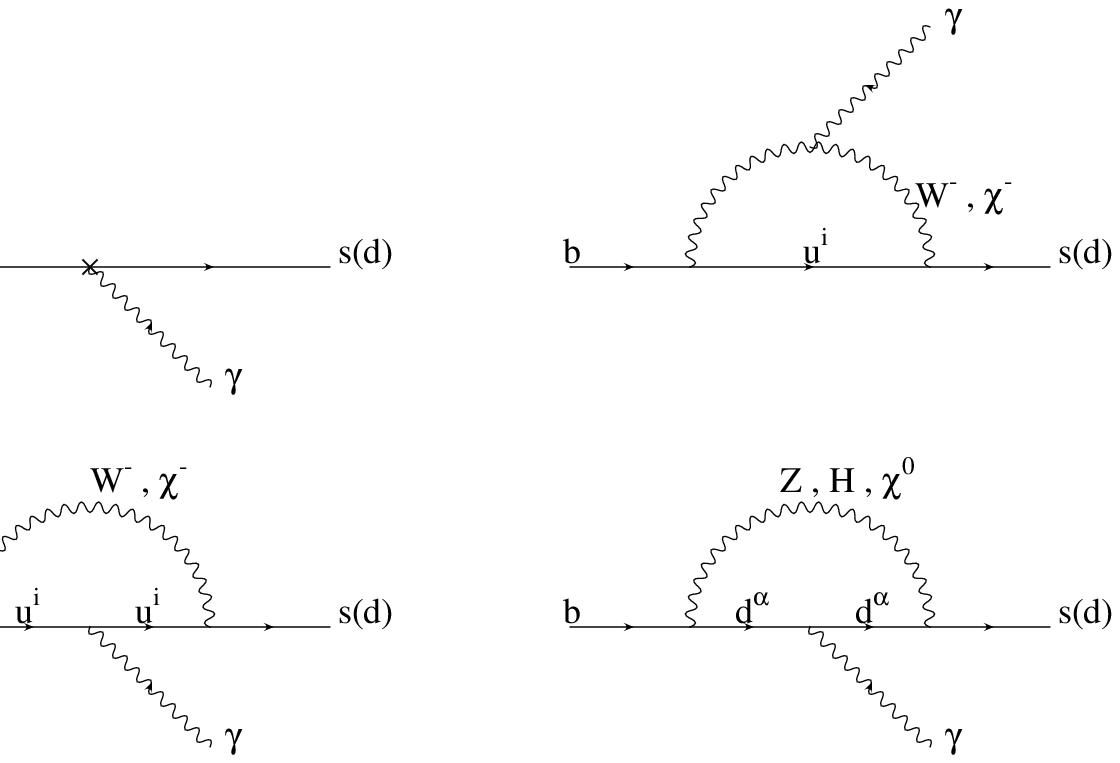}
	\caption{}
	\label{fig:one}
\end{figure}

Using on-shell renormalization, we just need to calculate the diagrams in Fig.
(\ref{fig:one}) only, and find the amplitude of on-shell
$b \rightarrow q \gamma$ up to second order in the external momenta as below,
\begin{eqnarray}
	T & = & \frac{G_F e}{8 \sqrt{2} \pi^2} \; \bar{q}(p^{\prime}) \;
		\left[\gamma_{\mu},\not{q}\right] \left(
		{F_L}^{\prime} m_q L + {F_R}^{\prime} m_b R \right) \; b(p) \;
		\epsilon^{\mu}
	\label{eqn:bsgdt}
\end{eqnarray}
where ${\xi_{\alpha}}^q \equiv {V_{CKM}^{\alpha q}}^{\ast} V_{CKM}^{\alpha b}$,
and
\begin{eqnarray}
	{F_L}^{\prime} & \equiv &
		Q_u \sum_{\alpha = u,c,t} {\xi_{\alpha}}^q F_1^{CC}(x_{\alpha})
		+ Q_d \left\{ \frac{2}{3} \sin^2\theta_W {z_d}^{qb}
		F_1^{NC}(r_q)\right. \nonumber \\
	& & \left. + \sum_{\alpha = d,s,b,4} {z_d}^{q \alpha} {z_d}^{\alpha b}
		\left[ F_2^{NC}(r_{\alpha},w_{\alpha})
		+ F_3^{NC}(r_{\alpha}) \right]
		\right\} , \label{eqn:fleft} \\
	{F_R}^{\prime} & \equiv &
		Q_u \sum_{\alpha = u,c,t} {\xi_{\alpha}}^q F_1^{CC}(x_{\alpha})
		+ Q_d \left\{ \frac{2}{3} \sin^2\theta_W {z_d}^{qb}
		F_1^{NC}(r_b)\right. \nonumber \\
	& & \left. + \sum_{\alpha = d,s,b,4} {z_d}^{q \alpha} {z_d}^{\alpha b}
		\left[ F_2^{NC}(r_{\alpha},w_{\alpha})
		+ F_3^{NC}(r_{\alpha}) \right]
		\right\} , \label{eqn:fright}
\end{eqnarray}
The suffixes $CC$ indicates the charged current contribution which also
appears in the SM (ref. \cite{inami}), and $NC$ indicates the neutral current
contribution which is charateristic of the model. The above functions are
given as,
\begin{eqnarray}
	F_1^{CC}(x_{\alpha}) & \equiv & x_{\alpha}
		\frac{-7 + 5 x_{\alpha} + 8 {x_{\alpha}}^2}
		{8 (1 - x_{\alpha})^3}
		- {3 x_{\alpha}}^2 \frac{(2 - 3 x_{\alpha}) \ln x_{\alpha}}
		{4(1 - x_{\alpha})^4} , \\
	F_1^{NC}(r_{\alpha}) & \equiv & \frac{-10 + 15 r_{\alpha}
		+ 6 {r_{\alpha}}^2 - 11 {r_{\alpha}}^3 - 6 r_{\alpha}
		\left( 3 - 4 r_{\alpha} \right) \ln r_{\alpha}}
		{6 (1 - r_{\alpha})^4} , \\
	F_2^{NC}(r_{\alpha},w_{\alpha}) & \equiv & r_{\alpha}
		\frac{-20 + 39 r_{\alpha} - {24 r_{\alpha}}^2
		+ 5 {r_{\alpha}}^3 - 6(2 - r_{\alpha}) \ln r_{\alpha}}
		{24 (1 - r_{\alpha})^4} \nonumber \\
	& &	- w_{\alpha} \frac{-16 + 45 w_{\alpha} - 36 {w_{\alpha}}^2
		+ 7 {w_{\alpha}}^3 - 6\left(2
		- 3 w_{\alpha}\right) \ln w_{\alpha}}
		{24 (1 - w_{\alpha})^4} , \\
	F_3^{NC}(r_{\alpha}) & \equiv & -\frac{-4 + 9 r_{\alpha}
		- 5 {r_{\alpha}}^3 - 6 r_{\alpha} \left( 1 - 2 r_{\alpha}
		\right) \ln r_{\alpha}}{12 (1 - r_{\alpha})^4} .
\end{eqnarray}
We denote $Q_u = 2/3$, $Q_d = -1/3$,
$x_{\alpha} \equiv {m_{\alpha}}^2/{{M_W}^2}$, $r_{\alpha} \equiv
{m_{\alpha}}^2/{{M_Z}^2}$ and $w_{\alpha} \equiv {m_{\alpha}}^2/{{M_H}^2}$.
Note that, $F_1^{CC}(x_t)$ is $W^{-}$ and $\chi^{-}$ exchange;
$F_1^{NC}(r_{\alpha})$ and $F_3^{NC}(r_{\alpha})$ are $Z$ exchange diagram
contribution. In $F_2^{NC}(r_{\alpha},w_{\alpha})$, the first term comes from
$\chi^0$ exchange, while the second term comes from $H$ exchange diagram.
In order to analyze the result further, we have to consider some constraints
on the mixings and the masses in the above equations. This will be the aim of
the next subsection.

\bigskip
\noindent
{\large \bf A. CONSTRAINTS FOR THE MIXING}

Before calculating the branching-ratio for $b \rightarrow s(d) \gamma$ in the
next subsection, let us impose
experimental constraints on the mixings which appear in eqs.
(\ref{eqn:fleft}) and (\ref{eqn:fright}). First, let us consider the mixing
${\xi_{\alpha}}^q$ in the charged current part. As discussed in refs.
\cite{morozumi} and \cite{nir}, in this model the unitarity of CKM matrix is
violated, and the relation
\begin{equation}
	{\xi_u}^q + {\xi_c}^q + {\xi_t}^q = {z_d}^{qb}  \label{eqn:triangle}
\end{equation}
is held. Nevertheless, since $m_u$ and $m_c$ are small compared with
$M_W$, the top-quark contribution in the $W^-$ and $\chi^-$
exchange diagrams is still dominant. Of course, this statement is valid
under the assumptions that $\left| {\xi_t}^q \right|$ is non-zero, i.e.
$\left| {\xi_t}^q/{\xi_u}^q \right| \gg 6 \times 10^{-9}$ and
$\left| {\xi_t}^q/{\xi_c}^q \right| \gg 5 \times 10^{-4}$.
So the up and charm quarks' contribution can not compete with the
top quark ones and we can neglect them. Thus, in the charged current part
it is sufficient to take only top quark	into account.

\begin{table}[t]
	\begin{center}
	\begin{tabular}{p{21mm}|c|c|l}
	Quark & Hadronic &  Related& Upper-bound \\
	subprocess & process & coupling & \\
 	\hline
	$b \rightarrow s \mu^+ \mu^-$ & $B \rightarrow X_s \mu^+ \mu^-$ &
	$\left| {z_d}^{sb} \right| $ & $\leq 1.7 \times 10^{-3}$ \\
	$b \rightarrow d \mu^+ \mu^-$ & $B \rightarrow X_d \mu^+ \mu^-$ &
	$\left| {z_d}^{db} \right| $ & $\leq 1.7 \times 10^{-3}$ \\
	$ s \rightarrow d \nu_e \bar{\nu}_e$ &
	$K^+ \rightarrow \pi^+ \nu \bar{\nu}$ &
	$\left| {z_d}^{sd} \right| $ & $\leq 5.9 \times 10^{-5}$ \\
	\end{tabular}
	\caption{Upper bound for $\left| {z_d}^{\alpha\beta} \right| $ from
	experiments, with assuming $Z$ exchange tree diagrams are dominant.}
	\label{tab:bound}
	\end{center}
\end{table}

On the other hand, for the mixing ${z_d}^{q \alpha} {z_d}^{\alpha b}$ in the
neutral current part, we find the upper-bounds on the related couplings
as written in Table \ref{tab:bound}, where we use the experiment results in
ref. \cite{data} and assume that the $Z$ mediated tree
diagrams are dominant. Using the upper-bounds in Table 1, we can neglect
the contribution of $\alpha = d$ for $q = s$ and $\alpha = s$ for $q = d$,
since $\left| {z_d}^{sd} {z_d}^{db} \right|$ and
$\left| {z_d}^{ds} {z_d}^{sb} \right|$ are tiny. Hence, the remaining mixings
are ${z_d}^{qq} {z_d}^{qb}$, ${z_d}^{qb} {z_d}^{bb}$ and
${z_d}^{q4} {z_d}^{4b}$. Furthermore, by using the definition of
${z_d}^{\alpha \beta}$ in eq. (\ref{eqn:w4}), ${z_d}^{q4} {z_d}^{4b}$
can also be written as,
\begin{eqnarray}
	{z_d}^{q4} {z_d}^{4b} & = & V^{q4} {V^{44}}^{\ast} \;
		V^{44} {V^{b4}}^{\ast} \nonumber \\
	& = & -{z_d}^{qb} \; \left| V^{44} \right|^2 .
\end{eqnarray}

\bigskip
\noindent
{\large \bf B. THE BRANCHING-RATIO}

In this subsection we will make a numerical calculation for
$b \rightarrow q \gamma$. Before going on, let us consider the
QCD correction up to leading logarithm in the model. Using the result in
ref. \cite{grig}, the QCD corrected amplitude is,
\begin{equation}
	T = \frac{G_F e}{8 \sqrt{2} \pi^2} \, Q_u {\xi_t}^q \;
		\bar{q}(p^{\prime}) \;
		\left[ \gamma_{\mu},\not{q}\right] \left(F_L^T(m_b) m_q L
		+ F_R^T(m_b) m_b R \right) \; b(p) \; \epsilon^{\mu},
	\label{eqn:dwqcd}
\end{equation}
where $F_L^T$ and $F_R^T$ are defined by,
\begin{eqnarray}
	F_L^T(m_b) & \cong & 0.68
		\left[ F_L - 0.42 F_L^g - 0.81 \right] , \\
	F_R^T(m_b) & \cong & 0.68
		\left[ F_R - 0.42 F_R^g - 0.81 \right] .
\end{eqnarray}
Here we suppose $\Lambda_{\rm QCD} = 150$(MeV) for $n_f = 5$, $\mu_0 = M_W$
and $\mu = m_b$. The functions $F_L^g$ and $F_R^g$
are induced by $b \rightarrow q g$ diagrams, which are realized by the third
and fourth diagrams in Fig. (\ref{fig:one}) with replacing the external photon
lines by the gluon ones. The functions are given as,
\begin{eqnarray}
	F_L & \equiv & F_1^{CC}(x_t)
		+ \frac{Q_d}{Q_u} \frac{{z_d}^{qb}}{{\xi_t}^q}
		\left[ \frac{2}{3} \sin^2\theta_W F_1^{NC}(r_q)
		\right. \nonumber \\
	& & \left. + {z_d}^{qq} \left( F_2^{NC}(r_q,w_q)
		+ F_3^{NC}(r_q) \right) + {z_d}^{bb} \left(
		F_2^{NC}(r_b,w_b)
		+ F_3^{NC}(r_b) \right) \right. \nonumber \\
	& & \left. - \left| V^{44} \right|^2 \left( F_2^{NC}(r_4,w_4)
		+ F_3^{NC}(r_4) \right) \right]  ,
		\label{eqn:flfin} \\
	F_R & \equiv & F_1^{CC}(x_t)
		+ \frac{Q_d}{Q_u} \frac{{z_d}^{qb}}{{\xi_t}^q}
		\left[ \frac{2}{3} \sin^2\theta_W F_1^{NC}(r_b)
		\right. \nonumber \\
	& & \left. + {z_d}^{qq} \left( F_2^{NC}(r_q,w_q)
		+ F_3^{NC}(r_q) \right) + {z_d}^{bb} \left(
		F_2^{NC}(r_b,w_b)
		+ F_3^{NC}(r_b) \right) \right. \nonumber \\
	& & \left. - \left| V^{44} \right|^2 \left( F_2^{NC}(r_4,w_4)
		+ F_3^{NC}(r_4) \right) \right]  ,
		\label{eqn:frfin} \\
	F_L^g & \equiv & F_2^{CC}(x_t)
		+ \frac{Q_d}{Q_u} \frac{{z_d}^{qb}}{{\xi_t}^q}
		\left[ \frac{2}{3} \sin^2\theta_W F_1^{NC}(r_q)
		\right. \nonumber \\
	& & \left. + {z_d}^{qq} \left( F_2^{NC}(r_q,w_q)
		+ F_3^{NC}(r_q) \right) + {z_d}^{bb} \left(
		F_2^{NC}(r_b,w_b)
		+ F_3^{NC}(r_b) \right) \right. \nonumber \\
	& & \left. - \left| V^{44} \right|^2 \left( F_2^{NC}(r_4,w_4)
		+ F_3^{NC}(r_4) \right) \right] ,
		\label{eqn:flgfin} \\
	F_R^g & \equiv & F_2^{CC}(x_t)
		+ \frac{Q_d}{Q_u} \frac{{z_d}^{qb}}{{\xi_t}^q}
		\left[ \frac{2}{3} \sin^2\theta_W F_1^{NC}(r_b)
		\right. \nonumber \\
	& & \left. + {z_d}^{qq} \left( F_2^{NC}(r_q,w_q)
		+ F_3^{NC}(r_q) \right) + {z_d}^{bb} \left(
		F_2^{NC}(r_b,w_b)
		+ F_3^{NC}(m_b) \right) \right.  \nonumber \\
	& & \left. - \left| V^{44} \right|^2 \left( F_2^{NC}(r_4,w_4)
		+ F_3^{NC}(r_4) \right) \right]  .
		\label{eqn:frgfin}
\end{eqnarray}
These equations are obtained by rewriting eqs. (\ref{eqn:fleft}) and
(\ref{eqn:fright}) with the constraints on the mixings which are derived
in the previous subsection. The charged current part in $F_L^g$ and $F_R^g$ is
given as (ref. \cite{buras}),
\begin{equation}
	F_2^{CC}(x_{\alpha}) = -x_{\alpha} \frac{2 + 5 x_{\alpha}
		- {x_{\alpha}}^2}{8(1 - x_{\alpha})^3}
		- \frac{3 {x_{\alpha}}^2 \ln x_{\alpha}}{4(1 - x_{\alpha})^4},
\end{equation}
From these equations, it is easily understood that there are three parts
in each equation, that is (1) the charged current, (2) the neutral current in
the ordinary quark sector and (3) the neutral current between the ordinary
quarks and the vector-like quark. Moreover, the quarks which appear in the
neutral current diagrams of $b \rightarrow q \gamma$ are
$d^{\alpha} = q, b, d^4$.

Next, let us evaluate each function in eqs. (\ref{eqn:flfin}),
(\ref{eqn:frfin}), (\ref{eqn:flgfin}) and (\ref{eqn:frgfin}).
According to the CDF's report (ref. \cite{cdf}), the mass of the top-quark
is $\sim 174 \pm 17$(GeV), so if we take the central value,
$F_1^{CC}(x_t) \sim -0.586$ and $F_2^{CC}(r_t,w_t) \sim -0.097$. Meanwhile, for
the light quarks (strangeness and down quarks), the value of the other
functions are nearly same, that is $F_1^{NC}(r_{light}) \sim -1.667$ and
$\left( F_2^{NC}(r_{light},w_{light}) + F_3^{NC}(r_{light}) \right)
\sim 0.333$.

By substituting these values, we obtain
\begin{eqnarray}
	F & \equiv & F_L \cong F_R \nonumber \\
	& \sim & -0.586 + \frac{Q_d}{Q_u}
		\left| \frac{{z_d}^{qb}}{{\xi_t}^q} \right| e^{i \theta}
		\left[ \frac{2}{3} \times 0.234 \times (-1.667)
		\right. \nonumber \\
	& & \left.
		+ \left( {z_d}^{qq} + {z_d}^{bb} \right) 0.333
		- \left| V^{44} \right|^2
		\left( F_2^{NC}(r_4,w_4) + F_3^{NC}(r_4) \right)
		\right] , \label{eqn:ffin} \\
	F^g & \equiv & F_L^g \cong F_R^g \nonumber \\
	& \sim & -0.097 + \frac{Q_d}{Q_u}
		\left| \frac{{z_d}^{qb}}{{\xi_t}^q} \right| e^{i \theta}
		\left[ \frac{2}{3} \times 0.234 \times (-1.667)
		\right. \nonumber \\
	& & \left.
		+ \left( {z_d}^{qq} + {z_d}^{bb} \right) 0.333
		- \left| V^{44} \right|^2
		\left( F_2^{NC}(r_4,w_4) + F_3^{NC}(r_4) \right)
		\right] , \label{eqn:fgfin}
\end{eqnarray}
where $ \theta \equiv {\rm arg} \left( {z_d}^{qb}/{\xi_t}^q \right)$.
The important point in eqs. (\ref{eqn:ffin}) and (\ref{eqn:fgfin}) is the
existence of the ratio $\left| {z_d}^{qb}/{\xi_t}^q \right|$ in the neutral
current part. The bounds on the ratio have been derived with eq.
(\ref{eqn:triangle}) and the experimental results for CKM matrix elements
as done in ref. \cite{morozumi},
\begin{eqnarray}
	\left| \frac{{z_d}^{sb}}{{\xi_t}^s} \right| & \le & 0.05 \: ,
	\label{eqn:bou1} \\
	\left| \frac{{z_d}^{db}}{{\xi_t}^d} \right| & \le & 2.15 \: .
	\label{eqn:bou2}
\end{eqnarray}
The bound on ${z_d}^{sb}$ (eq. (\ref{eqn:bou1})) tells us that the
contribution due to FCNC is smaller than that due to charged current.
Hence, we will here calculate the branching-ratio of $b \rightarrow d \gamma$
only.

In order to obtain the branching-ratio, we must know the dependence of
$F_2^{NC}$ and $F_3^{NC}$ on the vector-like quark mass $m_4$. In Fig.
(\ref{fig:two}), we depict $F_2^{NC} + F_3^{NC}$ as a function of $m_4$.
The figure tells us that the contribution due to the vector-like quark can
be comparable with the contribution due to top quark if
$\left| V^{44}\right|^2 \simeq O(1)$. In particular, when
$M_Z \le m_4 \le M_H$, the contribution is dominated by $F_2^{NC}$
(Fig. \ref{fig:three}). This explains why we must take the neutral Higgs
contribution $\left( F_2^{NC} \right)$ into account.
We have noticed that $F_2^{NC} \rightarrow 5/24$ when $r_{\alpha} \gg 1$
and $w_{\alpha} \ll 1$, i.e. the unphysical Higgs ($\chi^0$) contribution is
dominant and the physical Higgs ($H$) contribution is suppressed.
Note that the maximum value of $F_2^{NC}$ is about $0.17 \sim 0.18$
when $150 \le m_4 \le 250$(GeV) and $M_H \ge 750$(GeV).

Finally we obtain the branching ratio for the inclusive
$b \rightarrow d \gamma$ process,
\begin{eqnarray}
	Br(b \rightarrow d \gamma) & \cong &
		\frac{\alpha {G_F}^2}{128 \pi^4} {m_b}^5 \, \tau_B \, 
		{Q_u}^2 \left| {\xi_t}^d \right|^2 \; \left| F^T(m_b) \right|^2
		\nonumber \\
	& = &	\left( 1.105 \times 10^{-13} \right)
		\frac{1.5 \times 10^{-12}}{6.582 \times 10^{-25}}
		\; \left| {\xi_t}^d \right|^2 \; \left| F^T(m_b) \right|^2 ,
\end{eqnarray}
where we neglect the down quark mass $m_d$ because of ${m_d}^2 \ll {m_b}^2$,
and use the value $\tau_B = 1.5$(ps). The function $F^T(m_b)$ is defined as,
\begin{equation}
	F^T(m_b) \cong 0.68 \left( F - 0.42 F^g - 0.81 \right) ,
\end{equation}
with $F$ and $F^g$ are given in eqs. (\ref{eqn:ffin}) and (\ref{eqn:fgfin}).
At last, we present the branching-ratio as a function of $m_4$ and
the ratio $\left| {z_d}^{db}/{\xi_t}^d \right|$ with various values of $\theta$
in Figs. (\ref{fig:four}) and (\ref{fig:five}), with putting
$\left| V^{44} \right|^2 = \left| {z_d}^{dd} \right| = \left| {z_d}^{bb}
\right| \cong 1$ and letting $\left| {\xi_t}^d \right|$ as an unknown parameter.
Then, from Fig. (\ref{fig:four}) we obtain a theoretical region of the
branching-ratio in unit of $\left| {\xi_t}^d \right|^2$ as below,
\begin{equation}
	0.170 \le \frac{Br(b \rightarrow d \gamma)}
		{\left| {\xi_t}^d \right|^2} \le 0.262 \: .
\end{equation}
Moreover, the contribution of the vector-like quark is becoming constructive
when $0^0 \le \theta < 90^0$ and destructive when $90^0 < \theta \le 180^0$.

\subsection{\bf THEORETICAL STUDY ON THE SIZE OF FCNC AND THE QUARK MASSES}

\hspace*{1em} In the previous works (ref. \cite{morozumi}, \cite{nir},
\cite{branco}), it has been shown that the FCNC which is indicated by
${z_d}^{qb}$ is naturally suppressed by a factor $(m_q m_b)/{m_4}^2$.
If the statement is always valid, we will find
$\left| {z_d}^{db} \right| \le 6 \times 10^{-6}$ for $q = d$ with an
assumption that $m_4 \ge M_Z$. This leads to a consequence that the neutral
current contribution may compete with the charged current one only if
$\left| {\xi_t}^d \right|$ is the same order as that of
$\left| {z_d}^{db} \right|$. Here we point out that the FCNC even vanishes in
a special case. In order to show that, let us start with the following
down-type mass matrix
\begin{equation}
	M_d^0  = \left(
	\begin{array}{cccc}
		{m_d^0}	& 0	& 0	& 0 \\
		0	& {m_s^0}&  0 	& 0 \\
		0	& 0	& {m_b^0}& 0 \\
		 {J_d}	& {J_s}	&  {J_b}& {m_4^0}
	\end{array}
	\right) \: \: .
\end{equation}

It can be shown that with an appropriate choice of weak basis, we can always
transform the down-type mass matrix in this particular form. The necessary
condition to have non-zero ${z_d}^{qb}$ is the presence of $J_q$ and $J_b$.
We show that this is not sufficient condition here. To simplify the analysis,
suppose $J_s = 0$. At first sight, in this case, we may think that there is
FCNC between $d$ quark and $b$ quark because of the presence of $J_d$ and
$J_b$. However, if the original mass parameters $m_d^0$ and $m_b^0$ are the
same as each other, the FCNC between $d$ quark and $b$ quark vanishes.
To prove this, we need to diagonalize the following down-type mass matrix,
\begin{equation}
	M_d^0 {M_d^0}^{\dagger} = \left(
	\begin{array}{cccc}
		{m^0}^2	& 0	& 0	& m^0 J^{\ast} \\
		0	& {m_s^0}^2	& 0 	& 0 \\
		0	& 0	& {m^0}^2	& m^0 J^{\ast} \\
		m^0 J	& 0	& m^0 J	& 2 \left| J \right|^2 + {m_4^0}^2
	\end{array}
	\right) \: \: ,
	\label{eqn:mmd}
\end{equation}
where we assume that $J_d = J_b$ for the simplicity of the proof.
By assuming that $m^0 \ll \sqrt{2 \left| J \right|^2 + {m_4^0}^2}$, we find a
unitary matrix $V$ which diagonalizes the matrix (\ref{eqn:mmd}),
\begin{equation}
	V \cong \frac{1}{\sqrt{2}} \left(
	\begin{array}{cccc}
		-1 & 0 & -1 &
		\frac{2 m^0 J^{\ast}}{2 \left| J \right|^2 + {m_4^0}^2} \\
		& & & \\
		0 & \sqrt{2} & 0 & 0 \\
		& & &  \\
		1 & 0 & -1 & 0 \\
		& & &  \\
		\frac{\sqrt{2} m^0 J}{2 \left| J \right|^2 + {m_4^0}^2} & 0
		& \frac{\sqrt{2} m^0 J}{2 \left| J \right|^2 + {m_4^0}^2}
		& \sqrt{2}
	\end{array}
	\right) .
	\label{eqn:vmat}
\end{equation}
Under the above approximation, the mass eigenvalues are found by the
diagonalized $M_d^0 {M_d^0}^{\dagger}$,
\begin{eqnarray}
	M_d \, {M_d}^{\dagger} & \equiv &
		V \, M_d^0 \, {M_d^0}^{\dagger} \, V^{\dagger}
		\nonumber \\
	& \equiv & {\rm diag} \left({m_d}^2 \: , \: {m_s}^2 \: ,
		\: {m_b}^2 \: , \: {m_4}^2 \right)
		\nonumber \\
	& \cong & {\rm diag} \left(
		 \frac{{m^0}^2 {m_4^0}^2}{2 \left| J \right|^2 + {m_4^0}^2}
		 \: , \: {m_s^0}^2 \: , \: {m^0}^2 \: ,
		 \: 2 \left| J \right|^2 + {m_4^0}^2 \right) \: \: .
	\label{eqn:diam}
\end{eqnarray}
Here we denote $m_{\alpha}$ ($\alpha = d,s,b,4$) as the physical mass, while
$m_{\alpha}^0$ as the mass parameter in weak bases. There are two interesting
points in eq. (\ref{eqn:diam}) :
\begin{enumerate}
	\item Despite of assuming the same mass for down and bottom quarks
		first, we can derive the mass difference of the physical
		masses naturally. Note that $m_d$ is massless in the limit
		of vanishing the diagonal element $m_4^0$. This is because
		the rank of $M_d^0 \, {M_d^0}^{\dagger}$ is 3 in the limit.
	\item There is a constraint on the ratio of the mass mixing $J$ and
		the vector-like quark mass $m_4^0$, that is
		\begin{equation}
			\frac{2 \left| J \right|^2 + {m_4^0}^2}{{m_4^0}^2}
			\cong \frac{{m_b}^2}{{m_d}^2} \sim 2.5 \times 10^{5}
			\: . \label{eqn:mixing}
		\end{equation}
\end{enumerate}

Finally, we are now ready to evaluate the mixing ${z_d}^{db}$, ${z_d}^{b4}$
and ${z_d}^{d4}$. After substituting the above result (eqs. (\ref{eqn:vmat})
and (\ref{eqn:diam})) into eq. (\ref{eqn:w4}), we have
\begin{eqnarray}
	\left| {z_d}^{db} \right| & = & \left| V^{d4} {V^{b4}}^{\ast} \right|
		= 0 \: \: , \\
	\left| {z_d}^{b4} \right| & = & \left| V^{b4} {V^{44}}^{\ast} \right|
		= 0 \: \: , \\
	\left| {z_d}^{d4} \right| & = & \left| V^{d4} {V^{44}}^{\ast} \right|
		= \frac{\sqrt{2} m^0 J^{\ast}}{2 \left| J \right|^2 +
		{m_4^0}^2} \: \: .
\end{eqnarray}

Therefore, under the assumption which are considered, the mixing
${z_d}^{db}$ vanishes while the mixing between the lightest quark and the
vector-like quark, ${z_d}^{d4}$ is present. At this stage, it is worth-while
mentioning the mechanism of the vanishing of the FCNC. We can see that one of
the light quarks with the mass of $m^0$ decouples from the vector-like quark
by rotating the original weak basis with 45 degrees. This can be done only if
$m_d^0 = m_b^0$ in $M_d^0$. Therefore, there is no FCNC between $d$ quark and
$b$ quark, which leads to ${z_d}^{db} = 0$ in this case. The details of study
for various types of the mass matrix will be published elsewhere.

\subsection{\bf CONCLUSIONS}

\hspace*{1em} We have studied the most minimal extension of the SM, i.e.
introducing a down-type vector-like quark in addition to the SM, and showed
that there is a region where the neutral Higgs contribution is not negligible
and may be comparable with the $Z$'s one in the $b \rightarrow s(d) \gamma$.
Considering all of them, the contribution of the down-type vector-like quark
may be significant to the $b \rightarrow d \gamma$, but not to the
$b \rightarrow s \gamma$. In the last section, we argued the relation the relation between the FCNC and down-type quark mass matrix. We found that there is
a case when the FCNC among ordinary quarks vanishes while the FCNC between the
lightest quark and vector-like quark is present.

\bigskip
\noindent
{\large \bf ACKNOWLEDGMENTS}

We would like to thank T. Muta for comments and discussion about
the neutral Higgs contribution, and G.C. Branco for comments about
$ b \rightarrow s \gamma $ process. We also would like to thank
C.S. Lim, T. Inagaki, S. Hashimoto and
T. Yoshikawa for their useful discussion. This work is supported by the
Grant-in-Aid for Scientific Research ($\sharp 06740220$) from the Ministry of
Education, Science and Culture, Japan. The work of L.T.H. was supported by
a grant from the Overseas Fellowship Program (OFP-BPPT), Indonesia.

\noindent
{\it Note added} :

In Sec. IV of our original manuscript (Hiroshima Univ. preprint, HUPD-9409),
we have stated that FCNC between $d$ quark and $b$ quark is enhanced when
$m_d^0 = m_b^0$. This comment was wrong and we have corrected this point in
this new manuscript. The correct statement is that the FCNC between them
vanishes. We have corrected a part of abstract correspondingly.
Also in Table \ref{tab:bound}, we have corrected the value of
$\left| {z_d}^{sd}\right|$ and updated the values of
$\left| {z_d}^{db}\right|$ and $\left| {z_d}^{sb}\right|$.
Correspondingly, eqs. (\ref{eqn:bou1}), (\ref{eqn:bou2}) and
the Fig. (\ref{fig:five}) are corrected.

We also noticed a recent paper by G. Bhattacharyya, G.C. Branco and
D. Choudhury (ref. \cite{gautam}) whose the content overlapps with our
Sec. \ref{sec:proc}. However, in the paper the contribution of the neutral
Higgs and the dependence of the vector-like quark mass are not discussed. As
previously emphasized, one of the main point in Sec. III is that there is a
region where the neutral Higgs contribution can compete with the other ones.

\clearpage
\noindent
{\large \bf Figure Captions}
\begin{itemize}
	\item Fig. 1 \\
		Diagrams related to $b \rightarrow q \gamma$ ($q = s, d$) with
		down-type vector-like quark in on-shell renormalization. The
		counter term is realized by the $W^-$, $\chi^-$, $Z$, $H$ and
		$\chi^0$ exchange quark self-energy diagrams.
	\item Fig. 2 \\
		The value of the functions $F_2^{NC}$, $F_3^{NC}$ and
		$F_2^{NC} + F_3^{NC}$ as a function of $m_{\alpha}$, where
		$M_H \sim 750$(GeV). The dashed line is $F_2^{NC}$, the thin
		line is $F_3^{NC}$ and the thick line is $F_2^{NC} + F_3^{NC}$.
	\item Fig. 3 \\
		The value of the functions $F_2^{NC}$ as a function of
		$m_{\alpha}$ and $M_H$.
	\item Fig. 4 \\
		Branching-ratio of $b \rightarrow d \gamma $ as a function of
		$m_4$ with various values of $\theta$, where
		$m_t \cong 174$(GeV), $M_H = 750$(GeV),
		$\left| V^{44} \right|^2 = \left| {z_d}^{dd} \right| =
		\left| {z_d}^{bb} \right| = 1$ and
		$\left| {z_d}^{db}/{\xi_t}^d \right| = 0.93$. The dashed line
		is the case of VM with $\left| {z_d}^{db} \right| = 0$, the
		thick line and the thin line are the VM with non-zero
		$\left| {z_d}^{db} \right|$ and $\theta = 0^0, 180^0$.
	\item Fig. 5 \\
		Branching-ratio of $b \rightarrow d \gamma $ as a function of
		$\left| {z_d}^{db}/{\xi_t}^d \right|$ with various values of
		$\theta$, where $m_t \cong 174$(GeV), down-type vector-like
		quark mass $m_4 = 200$(GeV),
		$\left| V^{44} \right|^2 = \left| {z_d}^{dd} \right| =
		\left| {z_d}^{bb} \right| = 1$ and $M_H = 750$(GeV). The
		dashed line is the case of VM with
		$\left| {z_d}^{db} \right| = 0$,  the thick line and the thin
		line are the VM with non-zero $\left| {z_d}^{db} \right|$ and
		$\theta = 0^0, 180^0$.
\end{itemize}

\clearpage

\begin{figure}[h]
	\begin{center}
		\input{graph2.plt}
	\end{center}
	\caption{}
	\label{fig:two}
\end{figure}

\begin{figure}[h]
	\epsfysize=10cm
	\epsffile{graph3.eps}
	\caption{}
	\label{fig:three}
\end{figure}

\begin{figure}[h]
	\begin{center}
		\input{graph4.plt}
	\end{center}
	\caption{}
	\label{fig:four}
\end{figure}

\begin{figure}[h]
	\begin{center}
		\input{graph5.plt}
	\end{center}
	\caption{}
	\label{fig:five}
\end{figure}

\end{document}